\documentclass[aps,showpacs,superscriptaddress,groupedaddress]{revtex4}  
\usepackage{graphicx}  
\usepackage{dcolumn}   
\usepackage{upgreek}        
\usepackage{amssymb}   
\usepackage{blindtext}
\usepackage[T1]{fontenc}

\newcommand{\vecx}{{\bf x}}
\newcommand{\vecu}{{\bf u}}
\newcommand{\vecd}{{\bf d}}
\newcommand{\R}{{\mathbb R}}
\newcommand{\WT}{{\rm WT}}

\begin{document}

\widetext

\title{Heterogeneous bacterial swarms with mixed lengths}

\author{Shlomit Peled} \affiliation{Zuckerberg Institute for Water Research, The Jacob Blaustein Institutes for Desert Research, Ben-Gurion University of the Negev, Sede Boqer Campus 84990, Midreshet Ben-Gurion, Israel}

\author{Shawn D. Ryan} \affiliation{Department of Mathematics and Statistics, Cleveland State University, Cleveland, Ohio, USA} \affiliation{Center for Applied Data Analysis and Modeling, Cleveland State University, Cleveland, Ohio, USA}
\author{Sebastian Heidenreich} \affiliation{Department of Mathematical Modelling and Data Analysis, Physikalisch-Technische Bundesanstalt Braunschweig und Berlin, Abbestrasse 2-12, D-10587 Berlin, Germany}
\author{Markus B$\rm \ddot{a}$r} \affiliation{Department of Mathematical Modelling and Data Analysis, Physikalisch-Technische Bundesanstalt Braunschweig und Berlin, Abbestrasse 2-12, D-10587 Berlin, Germany}
\author{Gil Ariel} \affiliation{Department of Mathematics, Bar-Ilan University, Ramat Gan 52000, Israel}
\author{Avraham Be'er} \affiliation{Zuckerberg Institute for Water Research, The Jacob Blaustein Institutes for Desert Research, Ben-Gurion University of the Negev, Sede Boqer Campus 84990, Midreshet Ben-Gurion, Israel} \affiliation{Department of Physics, Ben-Gurion University of the Negev 84105, Beer-Sheva, Israel}

\vskip 0.25cm

\date{\today}

\begin{abstract}
Heterogeneous systems of active matter exhibit a range of complex emergent dynamical patterns. In particular, it is difficult to predict the properties of the mixed system based on its constituents. These considerations are particularly significant for understanding realistic bacterial swarms, which typically develop heterogeneities even when grown from a single cell. Here, mixed swarms of cells with different aspect ratios are studied both experimentally and in simulations. In contrast with previous theory, there is no macroscopic phase segregation. However, locally, long cells act as nucleation cites, around which aggregates of short, rapidly moving cells can form, resulting in enhanced swarming speeds. On the other hand, high fractions of long cells form a bottle-neck for efficient swarming. Our results suggest a new physical advantage for the spontaneous heterogeneity of bacterial swarm populations.
\end{abstract}

\maketitle

\section{Introduction}

In much of the natural environments, several species, or variants of the same species, occupy the same niches. Heterogeneity is an essential part of bacterial communities and in some cases, the different populations cannot be separated \cite{BenJacob2006,Blanchard2015,Finkelshtein2015,Hansen2007,Hibbing2010,Ingham2011,Kai2018,McCully2019,Nadell2016,Weber2014,Clement2016,VanGestel2015,VanDitmarsch2013,Deforet2019}. 
For example, colonies of {\it Bacillus subtilis} grown from a single cell spontaneously form two distinct sub-populations with strikingly different cell types, differing in motility, growth rate and cell size \cite{Kearns2005}. It has been hypothesized that the two cell-types have different "roles" in the colony; the majority of cells is smaller and highly motile, thus expanding rapidly to new territories, while a small fraction of long, less motile cells mostly exploit the present location. This point of view is principally biological in nature. Here, we take the first steps in studying heterogeneous bacterial swarms from physical perspectives, showing, both experimentally and in simulations, that mixed swarms may generate advantageous swarming conditions.

The vast work on bacterial swarming and collective bacterial swimming has focused on homogeneous systems with a single type of cell
\cite{Grobas2020,Jeanneret2019,Beer2019,Jeckel2019,Darnton2010,Sokolov2007,Ilkanaiv2017,Kearns2003,Kearns2010,Beer2020,Ariel2018,Huijing2012,Aboutaleb2017,Harshey2003,Copeland2009,Vallotton2013,Gachelin2013,Lopez2015}. Notable exceptions concentrate on the macroscopic growth of the colony and the spatial mixing (or lack of) between the populations \cite{Deforet2019,Zuo2020,Benisty2015,Partridge2018}. Motivated by the fact that bacterial cells change their aspect ratio prior to swarming \cite{Kearns2003,Kearns2010}, and that cell length dictates swarm dynamics \cite{Ilkanaiv2017,Beer2020}, in this paper we study experimentally the collective dynamics of mixed swarming bacterial populations composed of cells with different aspect-ratios, focusing on the microscopic swarming statistics, and compare the results to simulations of a physics-based model. According to the experimental phase diagram in \cite{Beer2020}, the dynamics of swarming wild-type (WT) cells is qualitatively different from the dynamics of elongated mutants (that differ in length only), showing distinct physical swarming regimes or phases. Accordingly, mixed colonies of such cells are particularly interesting, both from the biological perspectives and, as a highly studied example of active matter, from a physical point of view.

As theory and simulations of active matter establishes, heterogeneous systems of self-propelled agents show a range of interesting dynamics and a wealth of unique phases that depend on the properties of individuals. Examples include particles/agents with varying velocities \cite{Schweitzer1994,McCandlish2012,Mishra2012}, noise sensitivity \cite{Menzel2012,Ariel2015oren,Netzer2019}, sensitivity to external cues \cite{Book2017} and particle-to-particle interactions \cite{Copenhagen2016,Khodygo2019,Bera2020}. It was found that the effect of heterogeneity ranges from trivial (the mixed system is an average of two populations) to singular (one of the sub-populations dominates the dynamics of the group as a whole) \cite{Ariel2015oren}. Self-organizing emergent phenomena such as motility-induced spatial phase separation of spherical active particles \cite{Cates2015} and nonequilibrium clustering of self-propelled rods \cite{Bar2020} bear critical biological consequences on the colony and its ability to expand and survive \cite{Grafke2017,Zuo2020}. Thus, the physics of mixed swarms is of great significance to our understanding of realistic bacterial colonies. 

Previous modeling approaches of heterogeneous active matter or self-propelled particles have been used, with some levels of success, to study several aspects of mixed bacterial communities \cite{Blanchard2015,Kai2018,Kumar2014,Nambisan2020}. For example, on the macroscopic, colony-wide scale, continuous models of mixed bacterial colonies with different motility and growth rates show the balance between reproduction rates and the importance of moving towards the colony edge, where nutrients are abundant \cite{Deforet2019,Book2017}. Here, we implement an agent-based model, allowing us to test which cell-cell interactions give rise to the observed phenomena, facilitating a deeper probe into the physics underlying bacterial swarms.

\begin{figure}[ht]
\includegraphics[scale=0.8, trim=0 480 300 0, clip]{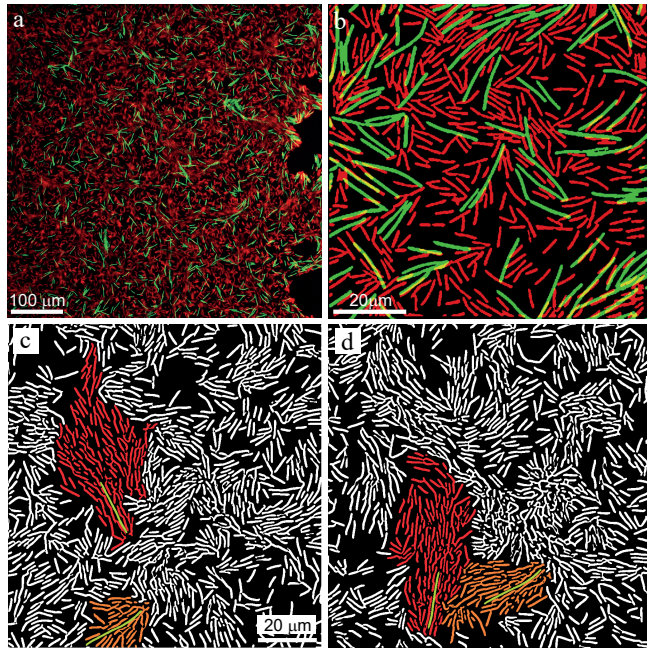}
\caption{\label{fig1} {\bf Experimental setup.}  A microscopic image of a mixed swarming colony; the strains were fluorescently labeled with different colors. The WT in red and the elongated cells (DS858) in green. (a-b) The mixed swarm at different magnifications. (c-d) A few elongated cells embedded in a WT colony. Two consecutive snapshots, 0.3 seconds apart, showing two elongated cells, each surrounded by a group of WT cells.}
\end{figure}

Our main finding is that introducing a small number of cells with a different aspect ratio than the majority can have a major effect on the dynamics of the swarm. To substantiate this claim, we present quantitative measurements of speed and density distributions in different mixtures. The cooperative action of many short cells mixed with a few longer cells leads to longer spatial correlations (indicating a more ordered swarming pattern) and higher average cell speeds. Figure~\ref{fig1} shows that a small number of long cells helps to organize the dynamics of the bacterial colony, with long cells acting as nucleation cites around which aggregates of short, rapidly moving cells can form. The impact of long cells is reproduced in a simple model based on hydrodynamic interactions, indicating a purely physical mechanism behind the beneficial effects of a few long cells on spatial organization and motion of all cells in the swarm. 

\section{Experiments}

Each experiment consisted of a 4-$\upmu$l drop of a heterogeneous bacterial culture, inoculated at the center of an agar plate (0.5\% agar and 25 g/l LB). Two {\it Bacillus subtilis} strains were used; the WT (strain 3610) with aspect ratio $7\pm2.2$ (mean $\pm$ standard deviation), and an elongated mutant (strain DS858) with a three-fold aspect ratio of $19 \pm 5.6$ \cite{Patrick2008}. See Fig. S1 \cite{supmat} for the length distribution. Mixing was done right prior to inoculation where each strain was grown separately overnight. Each of the strains was labeled fluorescently either green or red, forming two variants for each of the strains (green strain 4846 sfGFP, amyE::Pveg\textunderscore R0\textunderscore sfGFP\textunderscore spec, and red strain 4847 pAE1222-LacA-Pveg-R0\textunderscore mKate\#2 mls, amp). 
High resolution images were taken using an Optosplit II, Andor, hooked to a Zeiss Axio Imager Z2 microscope. The system splits the dually excited image (Ex 59026x, beam splitter 69008bs, and Em 535/30; 632/60) on a NEO camera ($1024 \times 2048$ and 50 frames/sec) in order to generate two simultaneous but separate fields of view, green and red, that were then merged again following post-processing using MATLAB.
See the Figs. S2-S4 \cite{supmat} and \cite{Guiziou2016} for details. We used a yellow set for imaging the green protein because the excitation light of the green set significantly affects the motility of the cells.  In all cases, the labeled cells behaved similarly to the non-labeled ones, with no photobleaching or reduction of speed due to intense illumination. Switching the colors of the long and WT cells from green to red showed no artifacts. Populations of the two strains were mixed at a variety of ratios prior to inoculation on the plates. Five hours after inoculation, each colony grew to a $\sim 5$cm in diameter disc with cells rigorously moving in a monolayer structure. We followed the outer regions of the colony where swarming is more pronounced. The statistics of the swarming dynamics is studied as a function of the total surface coverage $\rho$, and for several fractions of the elongated mutant (in terms of the surface coverage), denoted $f$ ($f=1$ only long cells, $f=0$ only WT cells). See for example Fig. S2, and Movie S1 \cite{supmat}.

\subsection{Experimental Results}

\begin{figure}[ht]
\includegraphics[scale=0.8, trim=0 490 300 0, clip]{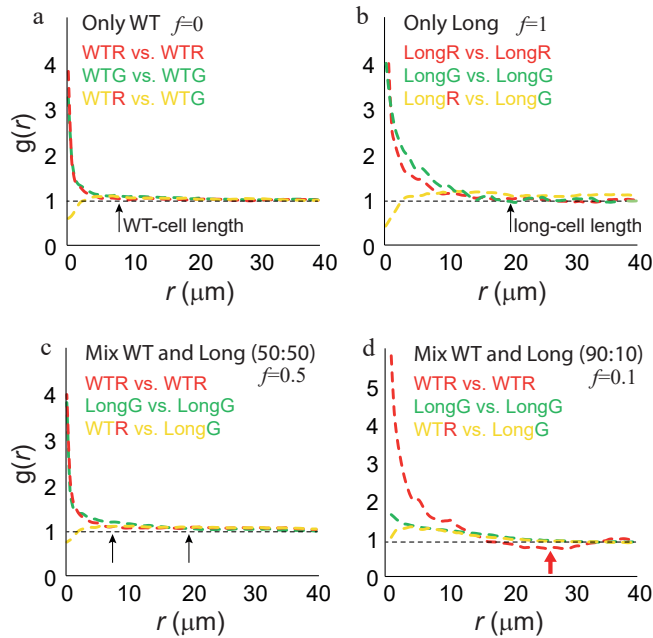}
\caption{\label{fig2} {\bf Experimental results – spatial correlations.} (a) The radial correlation function $g(r)$ for homogeneous WT populations: WT red variant vs. itself (red curve), WT green variant vs. itself (green curve), and WT red variant vs. WT green variant (yellow curve). (b) The radial correlation function for homogeneous populations of long cells: long cells red variant vs. itself (red curve), long cells green variant vs. itself (green curve), and long cells red variant vs. long cells green variant (yellow curve). (c) The radial correlation function in a mixed colony with $f=0.5$. WT vs. itself (red curve), long cells vs. itself (green curve), and WT vs. long cells (yellow curve). (d) The radial correlation function in a mixed colony with $f=0.1$. WT vs. itself (red curve), long cells vs. itself (green curve), WT vs. long cells (yellow curve).}
\end{figure}

On the macroscopic scale, the two strains formed well-mixed colonies, independently of the partial ratio of the sub-populations or the overall surface coverage. In particular, there is no phase segregation. Figure~\ref{fig2} shows the radial correlation function $g(r)$, which describes density variations in each sub-population for each of the strains (color labeled). Data was taken at $\rho \simeq 0.25$. Note that at very short distances ($<5\upmu{\rm m}$), our estimate for the radial distribution function, which was evaluated pixel content, is biased. See the appendix for definitions and estimation method. 
Figure~\ref{fig2}a shows results for a mix of red and green WT bacteria ($f=0$, same length, different colors); plots show the radial correlation function for each of the sub populations vs. themselves, and one sub population vs. the other one. Figure~\ref{fig2}b shows the same analysis for the elongated strain ($f=1$). These figures serve as control to verify that the fluorescence markers do not change the spatial distribution of cells. Figure~\ref{fig2}c shows results with a 50:50 mix of red WT and green elongated cells ($f=0.5$). In all cases, correlations beyond the size of a single cell were found to be negligible, indicating no long-range spatial correlation and a homogeneous distribution of cells (values remain close to 1 at distances larger than 40 $\mu$m – the maximum shown in Fig.~\ref{fig2}). In populations containing few long cells embedded in the majority of WT cells ($f=0.1$), the WT (shorter) cells align and partially aggregate around the long cells, which locally enhances order (Fig.~\ref{fig2}d). Interestingly, WT-WT correlations become negative (i.e., below 1) at length scales which are comparable to the length of long cells, indicating slight local attraction to long cells or formation of large clusters. Indeed, Figs.~\ref{fig1}c-d show a set of two snapshots with $\rho=0.38$ and $f=0.01$, taken 0.3 seconds apart. Tight groups of WT (short) cells clustering around a few long ones can be observed. Cells were colored manually to emphasize clusters.

\begin{figure}[ht]
\includegraphics[scale=0.8, trim=0 470 300 0, clip]{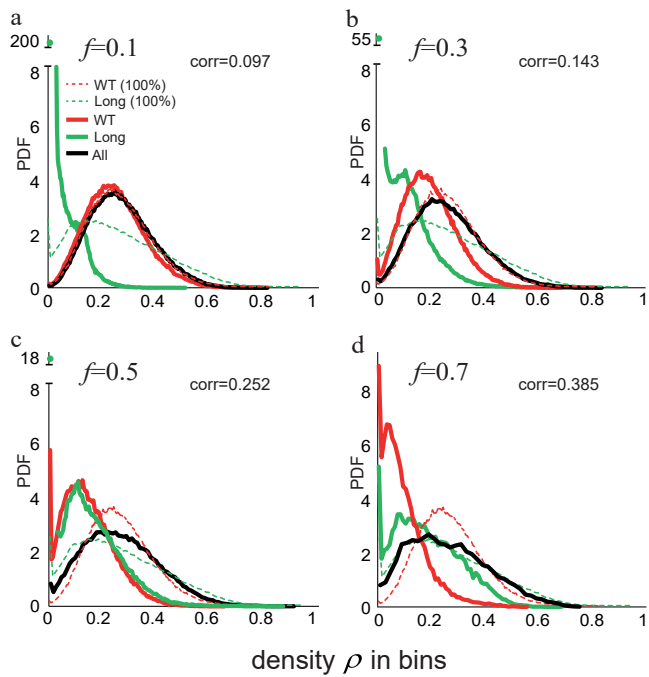}
\caption{\label{fig3} {\bf Experimental results – spatial distribution.} The distribution of WT (red), long (green) and total (black) cell density for fixed $\rho=0.25$ and different $f$, obtained by partitioning each image into $10 \time 10$ sub-domains. Dashed lines show the respective distribution of each cell type on its own ($f=0$ or 1). The spatial correlation between WT and long cells is noted in each subplot. The spatial distribution of long cells is qualitatively the same for all $f$, showing two local maxima – one at zero and one at 0.15-0.2. On the other hand, the distribution of WT cells is fundamentally different for small $f$ (<0.3) compared to larger $f$ values. For large and small $f$, the peak at 0 may be very large (e.g., $f=0.1$ green peak at 200; $f=0.3$ green peak at 55; $f=0.5$ green peak at 18).}
\end{figure}

The effect of one species on the dynamics of the other can also be observed in the spatial distribution of cells. Figure~\ref{fig3} shows the distribution of the surface coverage observed by partitioning the viewing field into $10 \time 10$ sub-domains at $\rho=0.25$ (calculated from the full images) for several example values of $f$. On their own, the spatial distribution of long cells ($f=1$) is wide, with two local maxima, one at 0 and the other one at ~0.2. The existence of empty sub-domains indicates a high degree of clustering. In contrast, WT cells on their own ($f=0$) are distributed more uniformly with the mode at around $\rho$. The effect of each bacterial species on the other is not symmetric. The spatial distribution of long cells always has two maxima, one at 0 and one around 0.15-0.2. Changing $f$, the height-ratio between the peaks varies. In contrast, the spatial distribution of WT cells changes qualitatively with $f$. For $f \ge 0.3$, a second maxima at zero emerges. Note that at $f=0.3$, and $\rho=0.25$, the surface fraction covered by long cells is $0.3 \cdot 0.25=0.075$, so the entire surface is covered by only 7.5\% long cells. Also, the fraction in number of cells is only 0.1 (or 10\%) because of their larger area (3 times larger compared to WT). This implies that a very small amount of long cells, both in number, and surface coverage, has a major effect on the spatial distribution of the entire swarm. Moreover, the location of the second maxima decreases with $f$. A possible explanation is that WT cells prefer to occupy regions with a small number of long ones. However, we find that the local density of WT and long cells is slightly positively correlated, typically around 0.1-0.4, with higher values at large $f$ (Fig.~\ref{fig3} for $\rho=0.25$).
This is consistent with the radial correlation functions depicted in Fig.~\ref{fig2}, showing that the inter-variant pair correlations is slightly higher compared to each type on its own.
In WT-WT experiments (a mix of red and green), the correlation between the density was found, as expected, to be 0. Therefore, the empty domains form because WT cells cluster close to long bacteria. 

\begin{figure}[ht]
\includegraphics[scale=0.8, trim=0 540 300 0, clip]{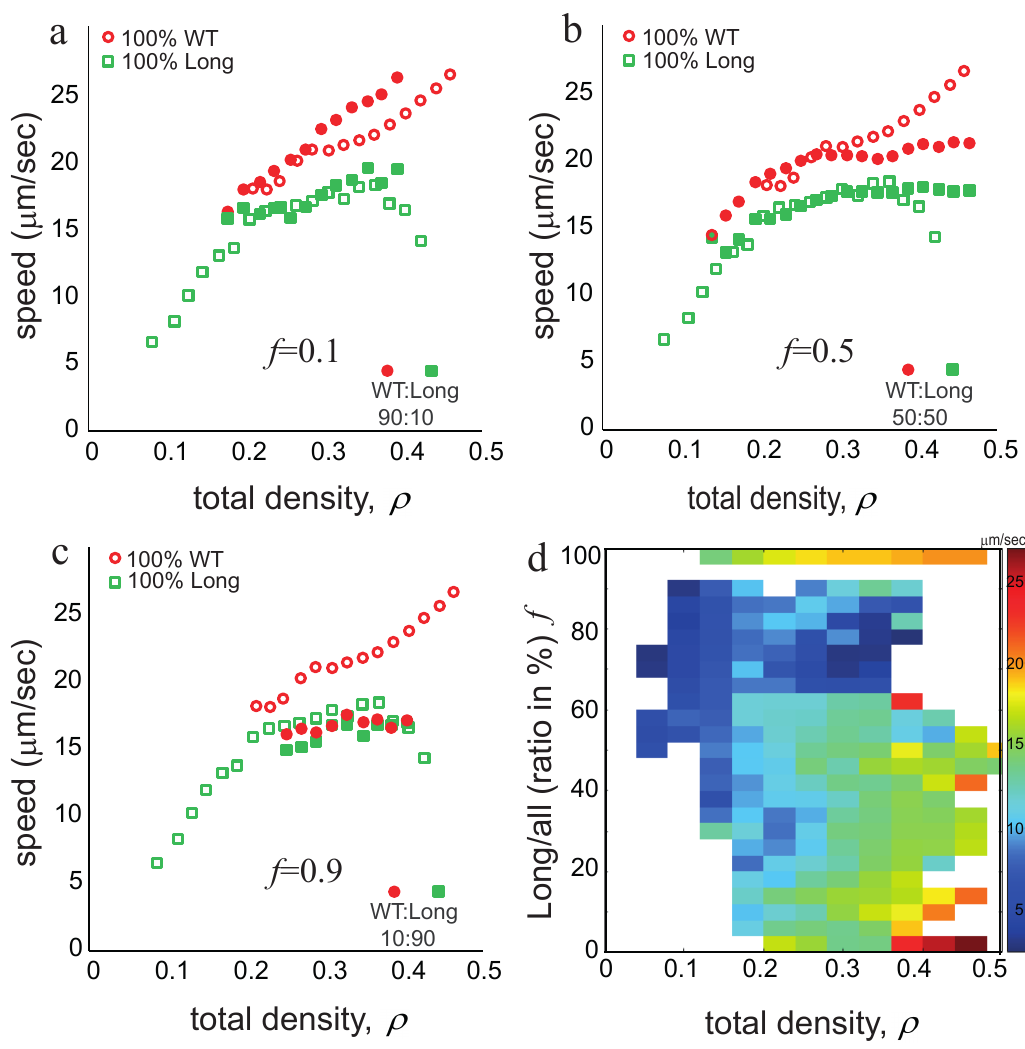}
\caption{\label{fig4} {\bf Experimental results – average microscopic speed.} Empty red circles, and empty green squares, show results for (a) Heterogeneous populations with a small fraction of elongated cells $f=0.1$. (b) Heterogeneous populations with an equal fraction of WT and elongated cells $f=0.5$. (c) Heterogeneous populations with a large fraction of elongated cells $f=0.9$. Solid red circles, and solid green squares, show results for the corresponding homogeneous systems. (d) The average speed as a function of the total surface coverage $\rho$, and the fraction of long cells $f$. Error bars are of the order of the circle/square size.}
\end{figure}

Next, we consider the dynamics of the heterogeneous colonies. Figure~\ref{fig4} shows the average speeds of each species for different $\rho$ and $f$, measured using standard optical flow algorithms. See \cite{Beer2019,Ilkanaiv2017} for details. The empty circles/squares in Figs.~\ref{fig4}a-c show the average microscopic speed of homogeneous systems (WT and elongated, i.e., $f=0$ or 1) as a function of $\rho$. WT cells move faster compared to elongated cells. Similar to past studies, the average speed increases with the total surface coverage \cite{Beer2020}. Figure S5 \cite{supmat} shows that specific labeling does not create an artifact in collective speed. In Fig.~\ref{fig4}a-c, the solid circles/squares show the average speed of heterogeneous systems, indicating the influence of one strain on the other. When the majority of the cells are WT, i.e., small $f$ (Fig.~\ref{fig4}a), elongated cells do not change their dynamics. However, the speed of WT cells is more complex. At high surface coverages (large $\rho$), WT cells move slightly faster compared to the case where they grow with no mixing. This effect can be explained by short (WT) cells aligning and aggregating around long cells (Fig.~\ref{fig1}c-d and Fig.~\ref{fig2}d), therefore, enhancing the local order and increasing the speed. The effect is reminiscent to drag reduction in turbulent flow \cite{Lumley1973}. In Fig.~\ref{fig4}b, with a 50:50 mixture ($f=0.5$), WT move slower (compared to $f=0.1$), while the speed of long cells is largely unaffected, except for a small increase at large $\rho$. When the majority of the cells are elongated, i.e., large $f$ (Fig.~\ref{fig4}c), the WT move dramatically slower, while the speed of long cells decreases slightly. To summarize this part, in Fig.~\ref{fig4}d we present a diagram showing that in general, the larger $\rho$ the faster the cells move, and that the larger the fraction of WT cells (smaller $f$) the faster the cells move. 

\section{Modeling and Simulation} 

In order to understand the interactions governing the dynamics of the mixed swarm, we suggest a simplified two-dimensional agent-based model of mixed two-species self-propelled particles (SPPs) with different aspect ratios. The main purpose of the model is to demonstrate that most of the experimental observations are of a physical origin. 

The model is derived from the balance of forces and torques on each cell. 
Several agent-based approaches have been suggested in the past, including dumbbells \cite{Hernandez2009}, regularized Stokeslets \cite{Cortez2005} and dipoles \cite{Ryan2011,Ryan2013,Ryan2016,Ariel2018,Ryan2019}. 
Here, we follow the approach of \cite{Ryan2011,Ryan2013,Ryan2016,Ariel2018,Ryan2019}, assuming a bacterium is essentially a point dipole where the size is incorporated through an excluded-volume potential and the shape is accounted for in the interaction of the point dipole’s orientation with the fluid. Tumbling is implemented as random turns in Poisson distributed times. This approach has been successfully used to study the dynamics of both motile and immotile cells \cite{Ryan2011,Ryan2013,Ryan2016,Ariel2018,Ryan2019}. One of the main advantages of the model is its computational simplicity, which is mostly due to the fact that the fluid equation for a point dipole has an exact analytical solution. In comparison to real swarming bacteria, the streamlines generated by the model in the intermediate to far-field are qualitatively similar, but the approximation breaks down at the cell surface. To make up for this inconsistency, a short-range purely repulsive potential is employed, acting as an effective excluded-volume interaction. 

The model assumes two types of particles: species 1, referred to as WT cells, have length $\alpha_{\rm WT}=7$, and species 2, referred to as long or elongated cells, have length $\alpha_{\rm long}=19$, $\alpha=\alpha_{\rm long}/\alpha_{\rm WT}$. In order to compare with experiments, the total surface coverage $\rho$ is calculated assuming minor axis width $b=1$. The simulation domain is a 2D square with periodic boundaries. 
Given the total surface coverage $\rho$ and the fraction of surface covered by long cells $f$, and assuming cells are ellipses, the number of WT cells, $N_{\rm WT}$, and the side length of the simulation domain $L$ are given by,
\begin{eqnarray}
   && \frac{\pi b}{4} \left[ \alpha_{\rm WT} N_{\rm WT} + \alpha_{\rm long} (N - N_{\rm WT}) \right] \nonumber \\
   && f = 1 - \frac{\pi b \alpha_{\rm WT} N_{\rm WT}}{4 \rho L^2} .
\nonumber
\end{eqnarray}
We assume that bacteria move in a thin film satisfying the Stokes equation and use the fluid velocity derived from two oppositely oriented force monopoles \cite{Cui2004}.
In \cite{Cui2004}, the authors assume that the thin film is held between two surfaces with no-slip conditions. 
In our case, this implies no-slip boundary conditions both on the agar surface and the
liquid-air interface. Indeed, previous works, e.g. \cite{Sokolov2007,Dunkel2013}, argue that in swarming {\it Bacillus subtilis}, the later interface should be modeled as no-slip due to secretion of surfactants and other bio-chemicals.
This is different than other systems, for example, dense swimming suspension of {\it E coli} that do not secrete surfactants, where  a free boundary condition is physically more realistic \cite{Chen2017}.

To be precise, the fluid velocity at position $\vecx \in \R^2$ and time $t$, $\vecu (\vecx ,t)$ is given by,
\begin{equation}
   \vecu (\vecx ,t ) = - \frac{l_1}{3 \pi} \sum_{i=1}^{N} \left( \nabla^3 \left[ \log ( | \vecx - \vecx_i | ) \right] \cdot \vecd_i (t) \right) p_i \vecd_i (t) ,
\label{eq:uxt}
\end{equation}
where $\vecx_i$ is the center-of-mass location, $\vecd_i$ is the (normalized) cell orientation, $p_i$ is the size of the cell dipole moment ($=p_\WT$ for WT and $p_{\rm long}$ for elongated cells) and $l_1$ is the film thickness. Assuming that the dipole moment is proportional to the length of the cell, we take $p_{\rm long}= \alpha p_\WT$.
Observe here that the flow generated at a point from either a short or long cell has the same basic form, but the magnitudes are different. In addition, noting that bacterial swimming results in a low Reynolds number flow, the cell dynamics are overdamped. Thus, considering the balance of forces and torques on each cell we have,
\begin{eqnarray}
   & m_i \dot{\vecx}_i =& v_0^i \vecd_i + \vecu ( \vecx_i,t) + \varphi \sum_{i \neq j} {\bf F} ( \vecx_i - \vecx_j ) 
 \label{eq:xdot} \\
   & m_i \dot{\vecd}_i =& - \vecd_i \times \left[ \nabla \times \vecu + \frac12 B_i \vecd_i \times ( \nabla \vecu + \nabla^T \vecu ) \vecd_i + \right. \nonumber \\
   && \left. \sum_{i \neq j} \omega^{\rm LJ} ( \vecx_i - \vecx_j ) \right] .
\label{eq:vdot}
\end{eqnarray}
Here, $m_i$ represents a dimensionless mass factor (=$m_\WT$ for WT and $m_{\rm long}$ for elongated cells, $m_{\rm long} = \alpha m_\WT$). 
As the dynamics is overdamped at low Reynolds number, this coefficient can also be understood as a friction constant.
The first term in Eq.~\ref{eq:xdot} represents self-propulsion of each bacterium in the direction it is oriented with (isolated) propulsion force $v_0^i$ (=$v_\WT$ for WT and $v_{\rm long}$ for elongated cells). Since the long cells have approximately the same flagellar density, the force excreted by the flagella is approximately proportional to $\alpha$, $v_{\rm long}=\alpha v_\WT$. 
As a result, the isolated swimming speed is the same for WT and elongated cells, as observed experimentally \cite{Beer2020}. 

The second term in Eq.~\ref{eq:xdot} describes the advection of the bacterium by the local flow $\vecu(\vecx,t)$, given by Eq.~\ref{eq:uxt}, generated by all the surrounding cells. The last term in Eq.~\ref{eq:xdot} is a short-range soft repulsion between cells. 
For simplicity, we use a truncated (purely repulsive) radial Lennard-Jones type potential that repels all cells within one cell length, $l$, and $\varphi$ represents the strength of this interaction which is the same for all cells (for additional details on the repulsive potential see \cite{Ryan2013}). 
Parameters for WT cells include the effective mass $m_\WT=1$, swimming speed $v_\WT=5$, and dipole moment $p_\WT=F_\WT \alpha_\WT b=\zeta  \eta a_\WT b v_\WT / m_\WT$, where the propulsion for $F_\WT$ is proportional to the isolated swimming speed $v_\WT / m$ and the ambient viscosity $\eta$ through an effective Stokes drag law for an ellipsoid with the drag coefficient $\zeta$ determined by the cell shape. 
In simulations, we take $\zeta \eta =1$. 
A more thorough analysis on the scaling of various coefficients on the aspect ratio may be pursued using slender-body theory \cite{Batchelor1970}. 
Indeed, assuming an ellipsoidal shape, suggests that some factors
that are logarithmic in the aspect ratio are neglected.
With the parameters corresponding to our experiment, the error introduced by this approximation is about 10-20\%, which is reasonable, considering the 
already much simplified physics that the dipole model assumes.

Eq.~\ref{eq:vdot} describes the time evolution of the orientation $\vecd_i$, which is given by Jeffery's equation. The first term represents the contribution to the orientational change due to the local vorticity while the second term in Eq.~\ref{eq:vdot} represents rotation due to the local shear. Assuming cells are prolate ellipsoids with aspect ratio $\alpha_i$, the shape is contained in the Bretherton constant $B=(\alpha_i^2-1)/(\alpha_i^2+1)$, which is between 0 (sphere) and 1 (pin). Hence, WT cells have $B=0.96$ while elongated cells have $B=0.99$.

Collisions between bacteria are accounted for in two ways. First, through a physical force from the excluded-volume potential described above. This forces the centers of mass of the colliding bacteria apart. In addition, when a collision occurs, i.e., when the distance between two particles is smaller than $b$, there is a realignment torque applied to the orientation represented by the last term in Eq.~\ref{eq:vdot}, where $\omega^{\rm LJ} (\vecx_i-\vecx_j )=|F^{\rm LJ} (\vecx_i-\vecx_j)|(\vecd^i \cdot \vecd^j)( \vecd^i \times \vecd^j)$. This term reinforces the following interaction between colliding bacteria: if particles collide with nearly parallel directions, then $\vecd^i \times \vecd^j$ is small. Similarly, if directions are nearly perpendicular, then $\vecd^i \cdot \vecd^j$ is small, resulting in a small torque. Also, the coefficient is the magnitude of the truncated LJ force ensures that this torque only is applied during collision events and does not contribute any other time to a bacterial motion (see \cite{Ryan2013} for further details).

In addition to the above, it has been shown that tumbling has an important effect on the resulting dynamics \cite{Ryan2016}. To this end, tumbling is modeled as a homogeneous Poisson process. When a cell is picked to tumble, a small rotational diffusion is added which is selected from a Gaussian distribution with zero mean $\mu=0$ and standard deviation $\sigma=\pi/12$. 

\begin{figure}[ht]
\includegraphics[scale=1.5, trim=0 685 10 0, clip]{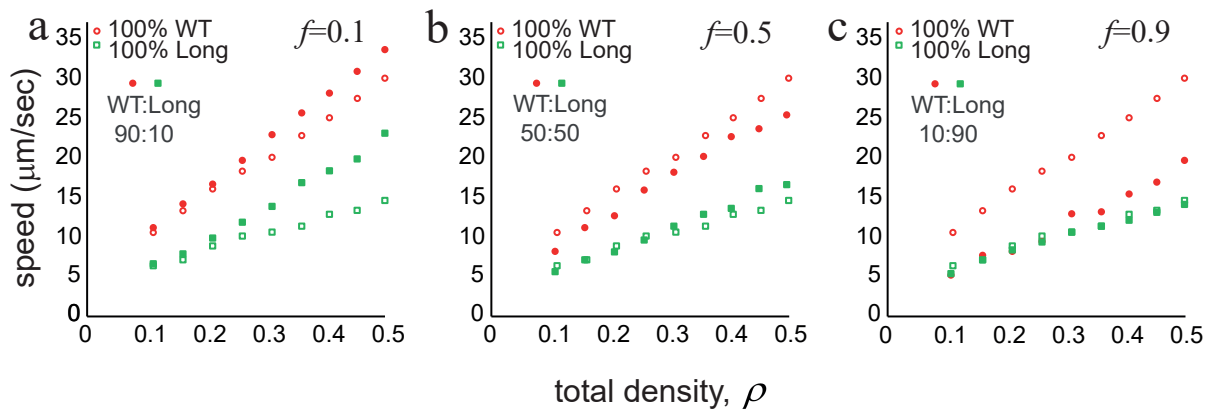}
\caption{\label{fig5} {\bf Simulation results.} (a-c) Average cell speed (solid symbols) as a function of surface coverage $\rho$. (a) $f=0.1$, (b) $f=0.5$ and (c) $f=0.9$. The empty symbols show simulation results for homogeneous systems – only WT (red) or only long (green). Compare with the experimental results shown in Fig.~\ref{fig4}. Error bars are of the order of the circle/square size.}
\end{figure}

In simulations, up to $N=2000$ cells are placed in a rectangular domain with periodic boundary conditions. Figure~\ref{fig5} shows that the dependence of the average particle speed on $f$ and $\rho$ agrees very well with experiments (Fig.~\ref{fig4}). First and foremost, the average speed of each species on its own increases with $\rho$. This result indicates that the hydrodynamic interaction increases order, enabling more efficient spreading. Experiments show that at high surface coverages the speed of long cells levels-off and even decreases slightly; simulations show a smaller decrease in speeds. This result is consistent with previous works showing that for long cells, steric effects, which are largely neglected by our model, become important at high densities \cite{Beer2020}.
Interestingly, tumbling has a non-negligible effect of the results (Fig. S6 \cite{supmat}).

The effect of adding a small number of long cells to a mostly WT swarm (small $f$) depends on the overall surface coverage. At small $\rho$, the average speed of both species is practically the same for each species on its own. This is reasonable, as the interaction between the species is weak. However, at larger surface coverages, both WT and long cells move faster. At larger $f$ this effect is not observed and the inherent speed of long cells seems to be a bottle-neck for efficient swarming. Similar to the experimental result, the speed of WT cells decreases with $f$ to match the speed of the longer bacteria. These effects are consistent with the experimental results (Fig.~\ref{fig4}). Finally, considering the spatial distribution of cells, simulations do not show statistically significant correlations between the densities of WT and long cells. This is expected taking into account the simplified cell-cell local interaction assumed by our model.

\section{Discussion} 

Our work presents the first experimental results describing the dynamics of heterogeneous bacterial swarms composed of cells with different aspect ratios. By controlling the overall surface coverage and ratio between the cell strains, we find new swarming regimes which are not expressed by the individual species on their own. In particular, we find that introduction of a small fraction of long cells ($f=0.3$, corresponding to about 10\% of the number of cells) fundamentally changes the spatial distribution of WT cells. At high surface coverages, the speed of WT cells depends non-monotonically on $f$. Introducing a small number of long cells to a WT swarm (low $f$) increases the average speed, while higher $f$ lowers the speed of WT cells, possibly due to multiple cell-cell collisions or jamming. 

Our experimental results contrast several theoretical predictions obtained in models of self-propelled rods and other types of active particles \cite{Kumar2014,Du2020,Nambisan2020,Winkler2020,Duman2018,Vliegenthart2020}. Most notably, although the speed of WT and long cells are different, there is no global phase separation and the swarm is well mixed on the macroscopic scale (in contrast to theoretical predictions for self-propelled rods \cite{Schweitzer1994,McCandlish2012,Mishra2012,Du2020,Zuo2020}. Locally, long cells increase clustering and the local density of WT and long cells are correlated. This result highlights the intricate interplay between short-range order and long-range mixing in active swarms.

Our model, consisting of self-propelled particles with hydrodynamic interactions, reproduces the speed dependence of both cell types at the entire range of $\rho$ and $f$ tested. However, the simulated spatial distributions of WT and long cells are not correlated. Therefore, hydrodynamic models of swarming bacteria also fall short at describing the full breadth of the dynamics. 
Additional experiments, including individual cell tracking may shed additional light on the correlations between the local orientational order and the coarse grained statistics such as average speeds and radial distribution function. 
Previous works with homogeneous swarms did not identify such correlations \cite{Beer2020}.
Extending these studies to mixed swarms is beyond the scope of the current manuscript.

Overall, our findings bring forth some of the intricate, non-intuitive biological aspects of realistic, heterogeneous swarms. In particular, we see that introducing even a small fraction of long cells, as occurs naturally in colonies of some swarming bacterial species, can significantly change the dynamics and the pattern of the group.

\section*{Acknowledgments}

We thank Daniel B. Kearns for sending the strains and Avigdor Eldar for creating the fluorescent variants. Partial support from The Israel Science Foundation’s Grant 373/16 and the Deutsche Forschungsgemeinschaft (The German Research Foundation DFG) Grant No. HE5995/3–1 and Grant No. BA1222/7–1 are thankfully acknowledged.

\section*{Appendix: radial correlation function}

Consider an ensemble of $n$ point particles at positions $\vecx_i$, $i=1,…,n$. 
The pair density $h(\vecx,\vecx')$ is defined as
\begin{equation}
   h(\vecx,\vecx') = \sum_{i \neq j} \delta (\vecx_i,\vecx) \delta (\vecx_j,\vecx) .
\nonumber
\end{equation}
Assuming the system is translation invariant, the pair correlation function is defined as
\begin{equation}
   g(\vecx) = \frac{1}{A n^2} \int \left< h(\vecx',\vecx'+\vecx) \right> d \vecx' ,
\nonumber
\end{equation}
where $A$ is the viewing area and $\left< \cdot \right>$ denotes an ensemble average. Assuming rotational invariance, the pair correlation function depends only on $r=|\vecx|$, and one can define the radial correlation function $g(r)=g(|\vecx|)$.
In experiments, the integral in $g(r)$ is approximated by sampling 10,000 points from 9 experimental images, taken approximately 1 second apart. For each point we average the red/green pixel content at distance r from the focal point. Thus, in this approximation, a particle is a red or green pixel, not a cell.


\end{document}